\documentclass[conference]{IEEEtran}
\usepackage[english]{babel}
\usepackage[T1]{fontenc}
\usepackage[latin1]{inputenc}
\usepackage{float}
\usepackage{graphicx}
\usepackage{amsmath,amssymb}
\usepackage{bm}
\usepackage{soul} 
\usepackage{color}
\usepackage{url}
\usepackage{balance}

\usepackage[acronym,shortcuts,nonumberlist, nopostdot,nogroupskip]{glossaries}
\makenoidxglossaries

\newacronym{5g}{5G}{fifth generation}
\newacronym{b5g}{B5G}{beyond-5G}
\newacronym{6g}{6G}{sixth generation}
\newacronym{5gc}{5GC}{5G core}
\newacronym{amf}{AMF}{access and mobility management function}
\newacronym{ausf}{AUSF}{authentication server function}
\newacronym{covid}{COVID-19}{coronavirus disease of 2019}
\newacronym{cran}{C-RAN}{cloud - radio access network}
\newacronym{csi}{CSI}{channel-state information}
\newacronym{crs}{CRS}{cell-specific reference signal}
\newacronym{d2d}{D2D}{device-to-device}
\newacronym{dmrs}{DMRS}{demodulation reference signal}
\newacronym{embb}{eMBB}{enhanced mobile broadband}
\newacronym{gnb}{gNB}{next-generation NodeB}
\newacronym{gdpr}{GDPR}{general data protection regulation}
\newacronym{gnss}{GNSS}{global navigation satellite system}
\newacronym{lte}{LTE}{long-term evolution}
\newacronym{ltea}{LTE-A}{long-term evolution-advanced}
\newacronym{lmf}{LMF}{location management function}
\newacronym{lbs}{LBS}{location-based service}
\newacronym{mec}{MEC}{multi-access edge computing}
\newacronym{mimo}{MIMO}{multiple-input multiple-output}
\newacronym{mmtc}{mMTC}{massive machine type communication}
\newacronym{mno}{MNO}{mobile network operator}
\newacronym{mptcp}{MP-TCP}{multipath transmission control protocol}
\newacronym{nef}{NEF}{network exposure function}
\newacronym{nwdaf}{NWDAF}{network data analytics function}
\newacronym{ngran}{NG-RAN}{next-generation radio access network}
\newacronym{nr}{NR}{New Radio}
\newacronym{ott}{OTT}{over-the-top}
\newacronym{pmi}{PMI}{precoding matrix indicator}
\newacronym{prs}{PRS}{position reference signals}
\newacronym{rat}{RAT}{radio access technology}
\newacronym{rfpm}{RFPM}{radio frequency pattern matching}
\newacronym{sba}{SBA}{service-based architecture}
\newacronym{smf}{SMF}{session management function}
\newacronym{srs}{SRS}{sounding reference signal}
\newacronym{tdoa}{TDoA}{time difference of arrival}
\newacronym{udr}{UDR}{unified data repository}
\newacronym{umts}{UMTS}{universal mobile telecommunications service}
\newacronym{ue}{UE}{user equipment}
\newacronym{upf}{UPF}{user plane function}
\newacronym{urllc}{URLLC}{ultra-reliable low-latency communication}
\newacronym{udm}{UDM}{unified data management}
\newacronym{vpmn}{VPMN}{virtual private mobile network}
\newacronym{wlan}{WLAN}{wireless local area network}

\newacronym{ip}{IP}{Internet protocol}
\newacronym{3gpp}{3GPP}{third-generation partnership project}

\begin{document}

%
\title{Virtual Private Mobile Network with Multiple Gateways for B5G Location Privacy}

\author{\IEEEauthorblockN{Stefano Tomasin and Javier German Luzon Hidalgo  }
	\IEEEauthorblockA{Department of Information Engineering\\
		University of Padova, Italy \\
		Email: tomasin@dei.unipd.it}
}


%


\maketitle

\begin{abstract}
 In a \ac{b5g} scenario, we consider a \ac{vpmn}, i.e., a set of \acp{ue} directly communicating in a \ac{d2d} fashion, and connected to the cellular network by multiple gateways. The purpose of the \ac{vpmn} is to hide the position of the \ac{vpmn} \acp{ue} to the \ac{mno}. We investigate the design and performance of packet routing inside the \ac{vpmn}. First, we note that the routing that maximizes the rate between the \ac{vpmn} and the cellular network leads to an unbalanced use of the gateways by each \ac{ue}. In turn, this reveals information on the location of the \ac{vpmn} \acp{ue}. Therefore, we derive a routing algorithm that maximizes the \ac{vpmn} rate, while imposing for each \ac{ue}  the  same data rate at each gateway, thus hiding the location of the \ac{ue}. We compare the performance of the resulting solution, assessing the location privacy achieved by the \ac{vpmn}, and considering both the case of single hop and multihop in the transmissions from the \acp{ue} to the gateways.
\end{abstract}


\IEEEpeerreviewmaketitle

\glsresetall

\section{Introduction}

The information on the position of smartphones and other connected devices is relevant for several location-based services, e.g. navigation, advertisement, and social networking. However, an uncontrolled disclosure of our position and movements is a severe violation of our privacy with  consequences at both personal and societal level, \cite{Gupta2017}. Some cases related to the disclosure of position have already been reported in courts, such as  covert location-based surveillance of employers, extra charges for rental car clients, circumstantial evidence gathering, and location-based profiling. Location privacy (also named geolocation privacy) has been studied from a legal standpoint in the last years \cite{king2011personal}, with a revamped interest associated to the recent \ac{gdpr} act in the European Union \cite{8484769}. Also these regulations have highlighted the responsibilities of the \acp{mno}, once considered as trusted.

The evolving cellular communication standard developed by the \ac{3gpp} has significantly strengthened the technical tools to locate and track \acp{ue}, with the \ac{5g} network being able not only to exploit millimeter waves for localization, but also artificial intelligence for its processing through the \ac{nwdaf}. Moreover, the \ac{mno} can disclose the position information to \ac{ott} operators through the \ac{nef}, further threatening our location privacy. The trend will continue in \ac{b5g} networks, where more precise localization techniques will be included, also using frequencies in the Terahertz (THz) band \cite{Fang20}.

\subsection{Related Literature}

Several works have addressed location privacy, in most cases focusing on network and application layers. A pretty good privacy solution has been proposed in \cite{DBLP:journals/corr/abs-2009-09035}, where connectivity and authentication functionalities were decoupled.  Other approaches are based on k-anonymity \cite{Li18} or differential privacy \cite{Yin18}. To limit the leakage of information to location-based service providers, a middle-ware can be introduced \cite{Beresford04}. Blockchains can also be exploited to anonymize users \cite{Qiu20},  satisfying the principle of k-anonymity privacy protection without the help of trusted third-party anonymizing servers. However, in all the existing literature the \ac{mno} has been assumed to be trusted and location privacy is defended only against external attackers. 

Only recently, a more global vision of location privacy protection has been introduced in \cite{1111111}, considering also the \ac{mno} as not trusted. In \cite{1111111}, the concept of \ac{vpmn} has been first introduced, where a set of \acp{ue} perform \ac{d2d} communications that are not accessible to the \ac{mno}, while only selected \ac{vpmn} devices operate as gateways between the \ac{vpmn} and the cellular network. With this approach, the \ac{mno} knows only the location of the gateways and   not that of the \acp{ue}.

\subsection{Contribution}

In this paper, we consider a \ac{vpmn} with multiple gateways, and  focus on the uplink transmission, where \ac{vpmn} \acp{ue} route their packets towards the gateways, which in turn forward them to the \ac{gnb} of the serving cell. \ac{ue} packets may arrive to the gateways either by a single hop, of passing though other \ac{vpmn} \acp{ue} operating as relays. In both scenarios, a first option is to apply a maximum-flow algorithm to determine the transmission rate. However, this choice creates an unbalance of data rates at the gateways, which partially reveals the location of the \ac{vpmn} \acp{ue}: in fact, the maximum rate solution uses close-by gateways more intensively. Therefore, we revise the maximum rate problem by adding the constraint that the same data rates are transferred by each gateway from each \ac{ue}. 

In \cite{1111111}, only very preliminary results on the performance of a possible \ac{vpmn} are reported, and here we offer a wider analysis of the \ac{vpmn}, in terms of probability of connectivity of the \acp{ue} in a given area, achievable rates, and localization error. Results show that the \ac{vpmn} offers an interesting trade-off among rates and  localization error in typical B5G cellular scenarios.

The rest of this paper is organized as follows. In Section~II, we revise the \ac{vpmn}, consider a possible implementation, and model the channels of the links. Section~III addresses the problem of routing for \ac{vpmn}, recalling the maximum-rate routing and introducing the location-privacy preserving routing. An in-depth evaluation of the considered \ac{vpmn} is proposed in Section~IV, before conclusions are driven in Section~V.

\section{The Virtual Private Mobile Network}

\begin{figure}
	\centering
	\includegraphics[width=1\hsize]{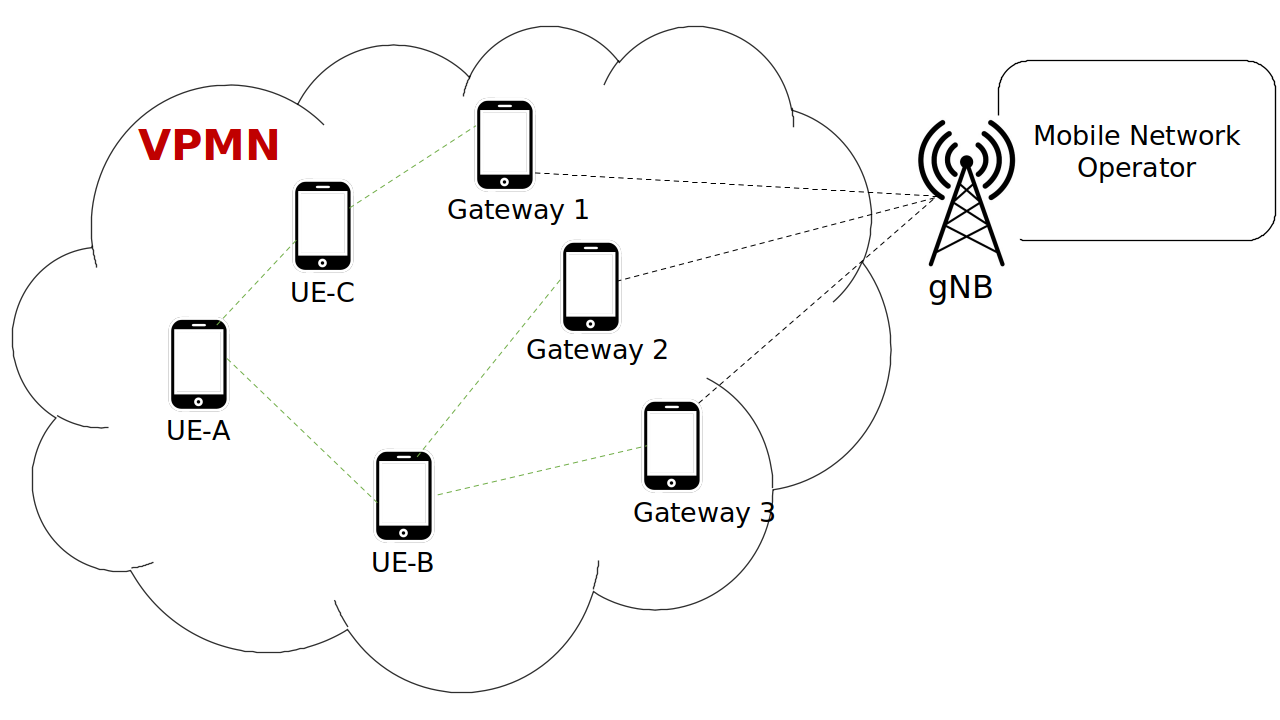}
	\caption{VPMN architecture with multiple gateways.}
	\label{fig:archvpmn}
\end{figure}

We focus here a \ac{vpmn} including $N$ devices, namely $M$ \acp{ue} and $S=N-M$ gateways, as introduced in \cite{1111111}. The objective is to prevent the localization of the \acp{ue} by the \ac{mno}. To this end, the \acp{ue} communicate in a \ac{d2d} fashion, without the intervention of the \ac{gnb}. The \ac{vpmn} will then communicate with the cellular network of the \ac{mno} through the gateways. An example of \ac{vpmn} is shown in Fig.~\ref{fig:archvpmn}. 

\subsection{Security Assumptions}

The \ac{mno} is only aware of the list of \acp{ue} in the \ac{vpmn} but it is not able to decode their communications or identify the signals over the air, i.e., associate them to the sender \ac{ue}. For this later aspect, suitable techniques should be deployed (see   \cite{1111111} and \cite{9171875}). In the uplink, when a \ac{vpmn} \ac{ue} has to send a packet to the cellular network, it transmits it, possibly by multiple hops within the \ac{vpmn}, to the  gateways, which in turn transmit it to the \ac{gnb}. Note that the communications among \acp{ue} will be encrypted with codes not available to the \ac{gnb}, to avoid localization of the \acp{ue} by the \ac{gnb}. In the downlink, the \ac{gnb} sends each packet in broadcast to the gateways, which will re-encrypt it and send to the destination \ac{ue} inside the \ac{vpmn}, possibly through multiple \ac{d2d} hops within the \ac{vpmn}. In the following, we focus on the uplink.

\subsection{VPMN Implementation}

Here we do not consider the implementation details of the \ac{vpmn}: we only observe that at the moment, the \ac{3gpp} standard does not allow its direct implementation. Indeed, the \ac{d2d} communication is strictly under the control of \ac{gnb} and devices involved in \ac{d2d} transmissions are localized and identified by the \ac{gnb}. We could implement the \ac{vpmn} as a femtocell, where the gateway  is the femto-\ac{gnb}: however, a single femto-\ac{gnb} would be available, restricting us to the use of a single gateway. 

A second option is to use another technology, e.g., the IEEE 802.11s standard that supports the implementation of a mesh WiFi network to implement the \ac{vpmn}, and the gateway \acp{ue} will interface the two networks. Although immediately deployable, this solution has the disadvantage of involving two standards, designed for two different scenarios; in particular, while cellular standards handle well mobility by the handover procedure, WiFi is designed for slowly moving terminals. In any case, we envision that a \ac{vpmn} can be well defined in future versions of the cellular standard, and this work is a contribution in this sense.

\subsection{Channel Model}

All \ac{vpmn} devices have a single antenna\footnote{Future studies will also consider devices with multiple antennas.}, and both the \ac{d2d} channels among \ac{vpmn} \acp{ue} and the gateway-\ac{gnb} channels are modeled including the effects of shadowing and path loss in an urban scenario. Two devices (\ac{ue} or gateways) will be connected if their channel gain value $\Gamma_{i,j}$ in dB satisfies 
\begin{equation}
\Gamma_{i,j} > \gamma,
\label{gammath}
\end{equation}
where $\gamma$ represents a suitable threshold in dB. Let $d_{i,j}$ be the distance between \acp{ue} $i$ and $j$.

The channel gain in dB is obtained from the addition of the local shadowing component at both the receiver ${z_j}$ and the transmitter ${z_i}$, and the path-loss  of the $i-j$ link  as
\begin{equation} \label{eq:1}
{{{\Gamma_{i,j}}}}=z_i + {z_j }+\log_{10}{{g_{i,j}}},
\end{equation}
where the path-loss of the $i-j$ link is 
\begin{equation} \label{eq:2}
{g_{i,j}}=\left(\frac{{d_{i,j}}}{r_0}\right)^{-\alpha},
\end{equation}
$r_0$ is a normalization factor, and $\alpha$ is the path-loss exponent. About the shadowing, we consider also its spatial correlation, and in particular the correlation of shadowing experienced at devices $i$ and $j$  is \cite{Book-Wiley}
\begin{equation} \label{eq:3}
{R_{i,j}}=\exp\left(-\frac{{d_{i,j}}}{d_{cor}}\ln{2}\right),
\end{equation}
where $d_{cor}$ is the decorrelation distance, dependent on the environment. We collect ${R_{i,j}}$ for all the devices into the $N\times N$ matrix $\bm{R}$. Then, let \bm{$z$} be a column vector collecting $\{z_i\}$, which can be generated as $\bm{z}=\bm{Lx}$, where $\bm{x}$ is a vector of independent zero-mean unitary variance Gaussian variables and $\bm{R}=\bm{LL}^H$ is  the Cholesky decomposition of  $\bm{R}$.

\section{Routing For The VPMN}

In the following, we will concentrate on the uplink, i.e., the transmission from the \ac{vpmn} \acp{ue} to the gateways and then the \ac{gnb}, as it can be exploited by the \ac{mno} to localize the \acp{ue}.

We consider two options for the routing of packets in the \ac{vpmn}: a) multihop, where packets generated by a \ac{ue} can be forwarded to a gateway by multiple hops though other \acp{ue}, and b) single hop, where \acp{ue} transmit packets directly to the gateways. About the single hop, we observe that a \ac{ue} that is not connected to any gateway is not part of the \ac{vpmn}. 

Time is divided into slots, and one \ac{ue} transmits in each slot to avoid interference. About the multihop case, the maximum achieved rate on link between devices $i$ and $j$ is 
\begin{equation}
\rho^{(\rm max)}_{i,j} = \begin{cases}
\log_2\left(1 +  \Gamma_{i,j}  \right),& \Gamma_{i,j} > \delta, \\
0, & {\rm otherwise,}
\end{cases}
\end{equation}
where we assumed without restriction a unitary-power additive white Gaussian noise. For the single-hop case we have 
\begin{equation}
\rho^{(\rm max)}_{i,j} = \begin{cases}
\log_2\left(1 +  \Gamma_{i,j}  \right),& j \in \{1, \ldots, S\} \mbox{ and  }\Gamma_{i,j} > \delta, \\
0, & {\rm otherwise},
\end{cases}
\end{equation}
where $1, \ldots, S$, are the indices of the $S$ \ac{vpmn} gateways and links with non-zero rates are only those towards the gateways.

\subsection{Maximum-Rate Uplink Routing Algorithm}\label{rout}
 
The routing solution that maximizes the achievable rate is the max-flow algorithm \cite{cormen2009introduction} on a graph having a node for each \ac{vpmn} device, and edges between all connected devices. The weight of the edge between devices $i$ and $j$ is its maximum rate $\rho^{(\rm max)}_{i,j}$. 

Note that the maximum flow problem is defined between one node operating as the source and one node operating as the sink, while in the case of multiple gateways we have multiple sinks. In this case, to obtain the maximum flow solution, we add a fictitious node, denoted as {\em super-sink}, which is connected to all gateways through edges with infinite weight (rate). Then, we solve the maximum flow problems from each \ac{ue} to the super-sink, using several solutions available, e.g., the Ford-Fulkerson algorithm \cite{cormen2009introduction}.

\subsection{Location Privacy Assessment}\label{locpres}

One main design objective for the \ac{vpmn} is to prevent the localization of \ac{vpmn} \acp{ue} by the \ac{mno}. Still, due to the fact that the \acp{ue} are connected (possible by multiple hops) to the gateways and the \ac{mno} can intercept packets going though the gateways, the \ac{mno} has some information on the \ac{ue} location. We now consider the localization error. 

In the following, we assume that all communications in the \ac{vpmn} are not identifiable and decodable by the \ac{mno}, i.e., the \ac{mno} cannot use them to obtain the location of \ac{vpmn} \acp{ue}.~\footnote{Note that this may require further attentions in designing communications among \acp{ue}, see \cite{1111111} for an overview of threats and possible countermeasures.} Still, the \ac{mno} knows the location of the gateways, using   several available localization techniques \cite{1111111}. Moreover, we assume that the \ac{mno} can intercept packets going through the gateways, and associate packets to its transmitting \ac{ue}, e.g., by the \ac{ip} address. Clearly, anonymization techniques operating at the \ac{ip} level can limit these possibilities.

To assess the localization error, let $\bm{p}_i$ be the position of \ac{ue} $i$, and let $\hat{\bm{p}}_i$ be its position estimated by the \ac{mno}.

\paragraph*{Localization Error With A Single Gateway} In case of a single gateway, the \ac{mno} sees all packets coming from a single position $\bm{p}_{\rm GW}$, that of the gateway.  Therefore, the best estimate of any \ac{vpmn} \ac{ue} position is the gateway position, i.e., $\hat{\bm{p}}_i = \bm{p}_{\rm GW}$. For the single hop scenario, the average localization error  
\begin{equation} \label{locerror}
\bar{U}  = \mathbb E[||\bm{p}_i - \hat{\bm{p}}_i||],
\end{equation}
coincides with the average distance of \acp{ue} from the gateway. Still, considering the correlated shadowing model, it is not straightforward to compute the average localization error. Therefore, in Section~\ref{numres} we will resort to simulations for its assessment.  

\paragraph*{Localization Error With Multiple Gateways} When multiple gateways are present, we can still estimate the average localization error as the average connection distance, resorting to numerical methods. However, we observe that when multiple gateways are present, the \ac{mno} can intercept packets and identify the transmitting \ac{ue}: by observing the data rates coming from the different gateways for a single \ac{ue}, and knowing the location of the gateways, a better localization can be achieved. For example, in the single hop scenario, if each \ac{vpmn} \ac{ue} is connected to the nearest gateway, observing from which gateway packets come, the \ac{mno} can significantly reduce the localization error. Similar considerations hold for the multihop scenario, where having multiple gateways allows also a trilateration based on the observed rates at each gateway. This will also be confirmed by simulations in Section~\ref{numres}. 

\subsection{Privacy Preserving Maximum Rate Routing}

We therefore propose  a routing strategy that ensures for each \ac{ue} that the same data rate is achieved through all the gateways, regardless of the \ac{ue} position. To this end, we formulate an optimization problem to determine the rate $f_{i,j}$ on link $i-j$.  First, the objective function to be maximized is the rate from \ac{ue} $v$ to all the gateways $s=1, \ldots, S$, i.e., (for all \acp{ue})
\begin{equation}\label{Cv}
  C(v) =  \sum_i \sum_{s=1}^{S} f_{i,s}.
\end{equation}
 Furthermore, to ensure that no information on the position of \ac{ue} $v$ is disclosed by the resulting flows through the gateways, we impose that all gateways have the same rate for each \ac{ue}, i.e.,
\begin{equation}
\sum_j f_{j,s_1} =  \sum_j f_{j,s_2} \quad \forall s_1, s_2 \in \{1, \ldots, S\}.
\label{samegat}
\end{equation}
We obtain the following linear programming problem:
\begin{subequations}
\begin{equation}
\max_{\{f_{i,s} \}} \quad     \sum_i \sum_{s=1}^S f_{i,s} 
\end{equation}
\begin{equation}
\textrm{s.t.} \quad    f_{i,j}\leq \rho^{(\rm max)}_{i,j} \quad \forall i,j \label{eq11}
\end{equation}
\begin{equation}
   \sum_i f_{k,i}\leq  \sum_j f_{j,k}   \quad \forall k > S, k \neq v   \label{eq12}
   \end{equation}
   \begin{equation}
 f_{i,j} \geq 0 \quad  \forall i,j\label{eq13} 
\end{equation}
\begin{equation}
\sum_j f_{j,s_1} =  \sum_j f_{j,s_2}  \quad \forall s_1, s_2 \in \{1, \ldots, S\}, \label{eq14}
\end{equation}
\end{subequations}
where \eqref{eq11} is the maximum flow constraint on the $i-j$ link, \eqref{eq12} is the flow-conservation constraint (the outgoing rate should not be larger than the incoming rate for each intermediate \ac{ue}), \eqref{eq13} ensures non-negative rates, while \eqref{eq14} is the constraint  \eqref{samegat} on the same rate for each gateway and each \ac{ue}.

\section{Numerical Results}\label{numres}

\begin{figure} 
	\centering
	\includegraphics[width=1\hsize]{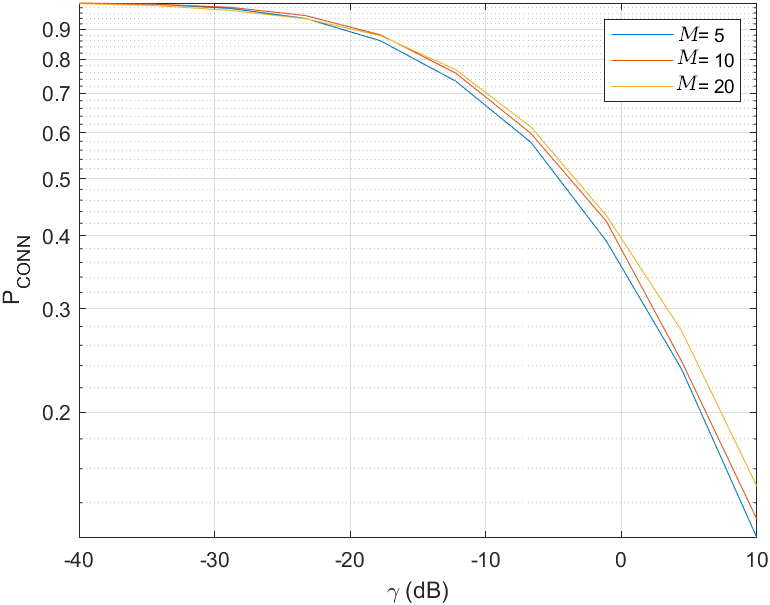}
	\caption{$P_\textrm{CONN}$ as a function of $\gamma$ for $N=5$, 10, and 20 devices, and multihop routing.}
	\label{fig:prob2}
\end{figure}

We consider an area of 100 m x 100 m, wherein $N$ \ac{vpmn} devices are uniformly randomly distributed. For channel modeling, we assume a decorrelation distance $d_{corr}= 20$~m and a path loss exponent $\alpha=2$. The path-loss normalization factor is $r_0=31.62$~m, corresponding to 10~dB loss every 100~m.  

We now assess the performance of the \ac{vpmn} according to several metrics. In particular, we evaluate the probability that all \acp{ue} constitute a single \ac{vpmn} (i.e., suitable connections with the gateways exist), the localization error by the \ac{mno} provided by the \ac{vpmn}, and the rates provided by the routing algorithms with and without location privacy constraint (achievable rates).

\subsection{Connectivity} \label{sec:1}

%
%
%

We consider here the probability that the $N$ randomly dropped devices constitute a \ac{vpmn} with multihop routing, i.e., they are a connected component where edges are present when condition \eqref{gammath} is satisfied.  

We recall that a {\em conforming set} $\mathcal C = \{(i,j)\}$ is a set of edges for which the $N$ nodes are connected. We denote with  $\Phi = \{\mathcal C\}$ the family of all conforming sets. The probability that all \acp{ue} are in the same connected component is
\begin{equation}
P_{\textrm{CONN}}=\sum_{\mathcal C \in \Phi} \mathbb P[\Gamma_{i,j}>\gamma, \forall (i,j)
\in \mathcal C,  \Gamma_{i,j}<\gamma, \forall (i,j) 
\not\in \mathcal C].
\end{equation}
Since $\Gamma_{i,j}$ are correlated Gaussian random variables, $P_{\textrm{CONN}}$ is related to the cumulative distribution of a set of correlated random variables, for which no close expression exists, thus we resort to numerical methods for its evaluation.

%
%
%
%
%

\begin{figure} 
	\centering
	\includegraphics[width=1\hsize]{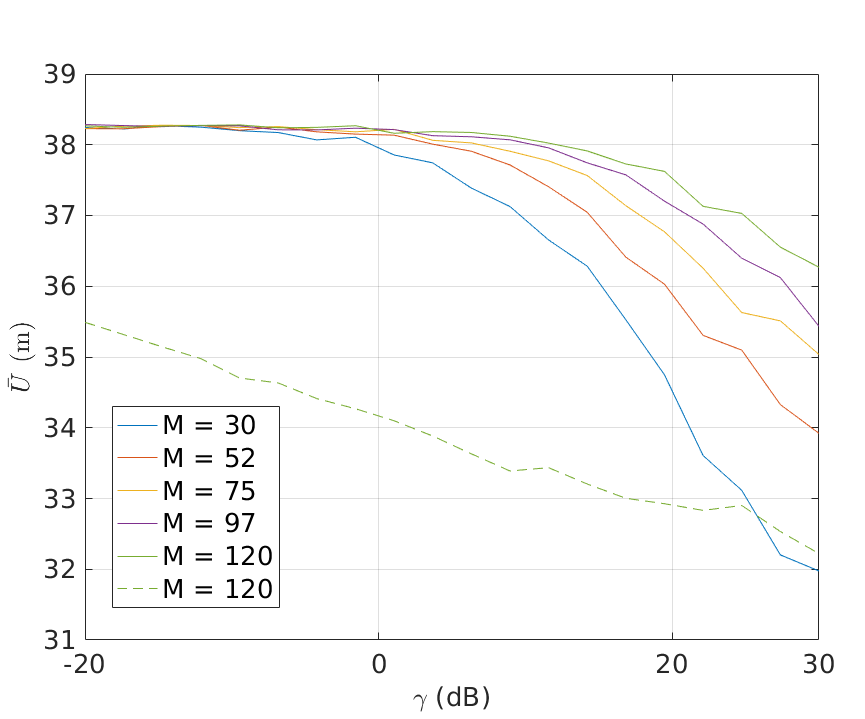}
	\caption{Average localization error  as a function of $\gamma$, for several values of $N$, with a single gateway: single hop (dashed line) and multihop (solid line) routing.}
	\label{fig:unc}
\end{figure}

Fig.~\ref{fig:prob2} shows the  probability that all devices are in the same connected component $P_{\rm CONN}$, as a function of $\gamma$ for $M=5$, 10, and 20 devices. We observe that increasing the number of devices  increases the connection probability: since the devices are randomly dropped in the same area, dropping more devices  will make them closer, thus increasing the probability of being connected. Moreover, decreasing $\gamma$ also increases the connection probability, with values above 0.9 for $\gamma = -20$~dB.

Note that $P_{\rm CONN}$ can be read also as the probability that a \ac{ue} finds a \ac{vpmn} to connect to, given that there are $N$ devices  in the area. Similar results hold also for the single-hop case, not reported here. 

\subsection{Localization Error}

We now consider the localization error, for the case of single and multiple gateways.

\paragraph*{Single Gateway} We consider a single gateway, placed at the center of our map  with coordinates $(0,0)$. Fig.~\ref{fig:unc} shows the average  localization error  $\bar{U}$ given by \eqref{locerror}, as a function of $\gamma$ for $M=30, 52, 75, 97$, and  $120$ \acp{ue}. We observe that for small values of $\gamma$, where we also have that all \acp{ue} are a connected component, the average localization error  is very high. For smaller connected components (higher values of $\gamma$) the localization error  is reduced.  For the multihop scenario instead, a higher number of nodes yields richer connected components, in turn increasing the localization error. In the single-hop scenario, the localization error does not depend on the number of \acp{ue}, as the connectivity of each \ac{ue} to the \ac{vpmn} does not depend on other \acp{ue}. Also, the single-hop scenario has a lower localization error  than the multihop case, due to the need of \acp{ue} to be directly connected to the gateways, i.e., being closer to each other.

\begin{figure} 
	\centering
	\includegraphics[width=1\hsize]{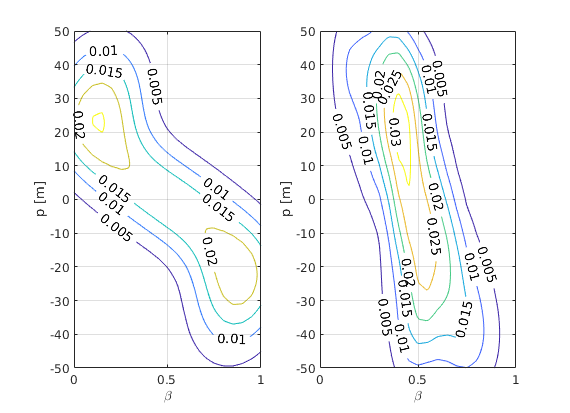}
	\caption{Contour plot of the estimated joint PDF of $(p,\beta)$, i.e., the \ac{ue} position and the flow ratio \eqref{defbeta} for the single hop (left) and multihop (right) scenarios.}
	\label{fig:probpos2}
\end{figure} 
 
\paragraph*{Multiple Gateways} When multiple gateways are present, we have observed that the \ac{mno} can infer the position of the \ac{ue} also by checking which fraction of the \ac{ue} traffic goes through the various gateways. To better understand this point, we consider a scenario where all devices are on a line: two gateways (with indices $1$ and $2$) are at positions $(0,20)$ and $(0,-20)$, while $M=28$ \acp{ue} are on the same line, with uniformly distributed positions in [-50, 50]. The ratio between the rates (solution of the max-flow problem) going through the two gateways is 
\begin{equation}
\beta = \frac{\sum_j f_{j,1}}{\sum_j f_{j,2}},
\label{defbeta}
\end{equation}
which is used to infer the position of the \ac{ue}. Fig.~\ref{fig:probpos2} shows the contour plot of the joint probability density function (PDF) $\phi(p, \beta)$ of the random vector $(p, \beta)$, with $p$ being the position of the \ac{ue}, and $\beta$ the corresponding value of the ratio \eqref{defbeta}. Both the single hop and multihop scenarios are considered. For a given observation of $\beta$, its maximum likelihood estimate of the \ac{ue} position is
\begin{equation}
\hat{p} = \mathbb E[p|\beta] = \int p \frac{\phi(p, \beta)}{\phi(\beta)} dp\,,
\end{equation}
where $\phi(\beta)$ is the PDF of $\beta$, which can be obtained from $\phi(p, \beta)$ by marginalization. With this procedure, the average localization errors   for the single hop and multihop scenarios are
\begin{equation}
\bar{U}_{\rm SH} =  3.9~{\rm m}, \mbox{ and } \bar{U}_{\rm MH} = 26~{\rm m}.
\end{equation}
Clearly, in the single hop scenario the PDF of Fig.~\ref{fig:probpos2} is much more concentrated (with smaller variance) than the PDF in a multihop scenario, since for the single hop scenario it only depends on the direct links between the \ac{ue} and the gateways, which in turn mostly depend on the distance of the node from the gateways through the path-loss.

\subsection{Achievable Rate}

\begin{figure}
	\centering
	\includegraphics[width=1\hsize]{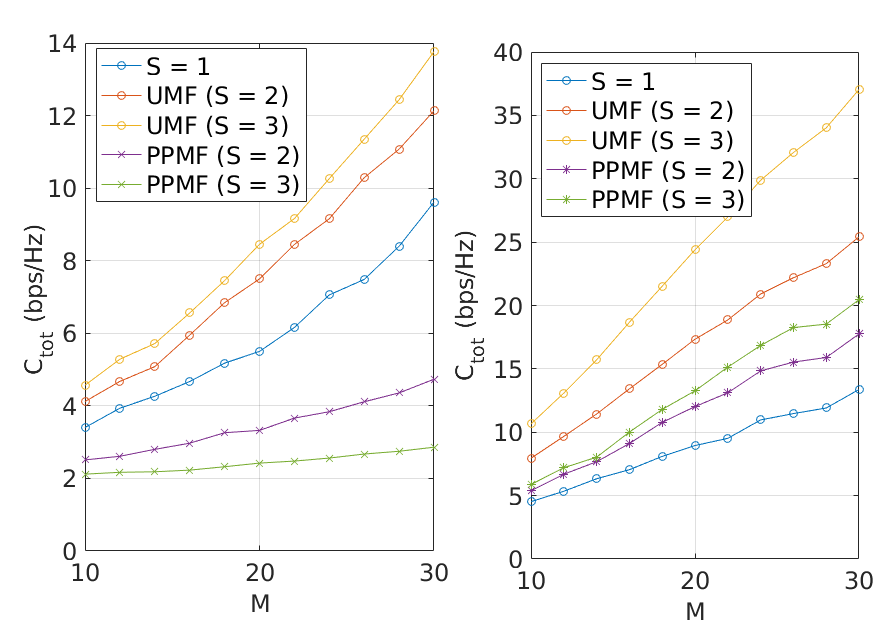}
	\caption{Average rate as function of the number of \ac{vpmn} devices for $S=1$ to 3 gateways, for the single-hop (left) and multihop (right) routing, using unconstrained max-flow (UMF) and privacy-preserving max-flow (PPMF) routing.}
	\label{fig:maxflow3}
\end{figure}

We now consider the average rate through the gateways, namely, the average of the sum of \eqref{Cv} over all \acp{ue} 
\begin{equation}
C_{\rm tot} = \mathbb E\left[\sum_v C(v)\right],
\end{equation}
where the average is done with respect to the \acp{ue} positions and channel realizations for randomly dropped devices in the square area. Fig.~\ref{fig:maxflow3} shows the average rate  for $S= 1, 2$, and 3 gateways, and different numbers of \acp{ue}, $M$. We compare the performance obtained with the unconstrained maximum-flow (UMF) algorithm and the privacy-preserving maximum flow (PPMF) solution of Section~\ref{locpres}. First, note that with a single gateway the UMF and PPMF solutions coincide, thus we show a single line per scenario. Then, we note that the single-hop scenario yields much lower rates than the multihop scenarios, where links among \acp{ue} are exploited to increase the rate. Lastly, we observe that introducing the privacy preserving constraints has a significant impact on the rate, in particular for the single-hop scenario, where the direct links of the \acp{ue} with all the gateways may not be available, and in this case the rate is zero to avoid location information disclosure. In the multihop scenario, instead, the rate reduction is much less pronounced, making this option more attractive for implementation.

%
%
%
%
%

\balance

\section{Conclusions}

For a \ac{vpmn} aiming at defending the location privacy of the \acp{ue} while ensuring connectivity with the cellular network, we have considered the case of multiple gateways. We have highlighted that using an internal routing algorithm that simply maximizes the \ac{vpmn} data rate towards the cellular network may reveal information on the location of the \ac{vpmn} \acp{ue}. Therefore, we have derived a routing solution that prevents this information leakage by ensuring that data is collected from each \ac{ue} with the same rate from all the gateways. We have also assessed the performance of the proposed routing solution, showing that it has a reduced performance loss with respect to the maximum-rate routing.

\bibliographystyle{ieeetr} 
\bibliography{template-thesis}

\end{document}